\documentclass{aa}
\usepackage{graphicx}
\usepackage{txfonts}
\usepackage{hyperref}
\usepackage{ulem}
\usepackage{float}
\usepackage{comment}
\usepackage{caption}
\hypersetup{
    colorlinks=true,
    linkcolor=blue,
    citecolor=blue,
    filecolor=magenta,      
    urlcolor=cyan,
    pdftitle={Overleaf Example},
    pdfpagemode=FullScreen,
    }

\newcommand{\Lya}{Ly$\rm{\alpha}$\,}

\newcommand{\Hb}{H$\rm{\beta}$ }

\newcommand{\OIII}{[\ion{O}{III}]}

\newcommand{\pc}{SMACS-PC-z7p7}
\newcommand{\jwst}{{\it JWST}}

\begin{document} 

   \title{Not all protoclusters host evolved galaxies: Evidence for reduced environmental effects in a lower halo mass protocluster at $z =7.66$}
   \titlerunning{Reduced environmental effects in a $z=7.66$ protocluster}

   \author{Callum Witten\inst{1}    
    \and Pascal A. Oesch\inst{1, 2, 3}
    \and Jake S. Bennett \inst{4}
    \and Romain A. Meyer \inst{1}
    \and Emma Giovinazzo \inst{1}
    \and Alba Covelo-Paz \inst{1}
    \and William M. Baker \inst{5}
    \and Lucy R. Ivey \inst{6,7}
    }
    
   \institute{Department of Astronomy, University of Geneva, Chemin Pegasi 51, 1290 Versoix, Switzerland
   \and
   Cosmic Dawn Center (DAWN), Copenhagen, Denmark
   \and
   Niels Bohr Institute, University of Copenhagen, Jagtvej 128, 2200 Copenhagen, Denmark
   \and
   Center for Astrophysics | Harvard $\&$ Smithsonian, 60 Garden Street, Cambridge, MA 02138, USA
   \and 
   DARK, Niels Bohr Institute, University of Copenhagen, Jagtvej 155A, DK-2200 Copenhagen, Denmark
   \and 
   Kavli Institute for Cosmology, University of Cambridge, Madingley Road, Cambridge, CB3 0HA, UK
   \and 
   Cavendish Laboratory, University of Cambridge, 19 JJ Thomson Avenue, Cambridge CB3 0HE, UK
   \\
   \email{callum.witten@unige.ch}
   }

   \date{}
 
  \abstract 
    {The progenitors of present-day galaxy clusters offer crucial insight into how galaxies and large-scale structure co-evolve in the early Universe. We present \jwst /NIRCam grism spectroscopy of the photometrically identified $z=7.66$ protocluster core in the SMACS J0723.3-7327 lensing field, \pc. We find six [\ion{O}{III}]-emitters and five additional photometric candidates within a 0.3 arcmin$^2$ ($1.5\ {\rm cMpc}^2$) region, corresponding to an overdensity of $\delta \sim 200$. Despite the extreme overdensity, the resident galaxies exhibit star formation histories, UV slopes, and neutral hydrogen column densities that are consistent with those of field galaxies at similar redshifts. This is in stark contrast to the consistently high neutral hydrogen column densities, old stellar populations, and large dust masses of galaxies within a $z=7.88$ protocluster in the Abell 2744 field. Comparison with the TNG-Cluster and TNG300 simulations indicates a halo mass of ${\rm log_{10}}(M_{200{\rm c}}[{\rm M_{\odot}}]) = 11.4\pm0.2$ and implies that, on average, \pc\ will evolve into a present-day Fornax-like cluster (${\rm log_{10}}(M_{200{\rm c},\ z=0}[{\rm M_{\odot}}]) = 14.3\pm0.6$). The uniformly young, highly star-forming nature of the galaxy population of \pc\ suggests that environmental effects only become significant above halo masses of ${\rm log_{10}}(M_{200{\rm c}}[{\rm M_{\odot}}]) \gtrsim 11.5$. Comparison with other $z\gtrsim7$ protoclusters reveals that vigorous star formation persists in lower-mass protoclusters, whereas accelerated evolution and suppression of star formation emerge in more massive haloes. \pc\ therefore represents an early stage of protocluster assembly, in which residence within an overdense environment still enhances star formation and feedback processes have yet to exert a significant influence.}
   \maketitle

\section{Introduction}
In the $\Lambda$ cold dark matter \citep[$\Lambda$CDM;][]{peebles_large-scale_1982, blumenthal_formation_1984} model of the Universe, the dark matter (DM) distribution is inhomogeneous on small scales. Baryonic matter falls onto early DM haloes, forming the first stars and galaxies, while on larger scales ($\sim 0.1 - 1$ pMpc) DM overdensities result in loose collections of galaxies. With cosmic time, these regions become virialised. At a transitional DM halo mass (${\rm log_{10}}(M_{\rm Halo}\ \left[{\rm M_{\odot}}\right]) \sim 11.8$; \citealt{dekel_galaxy_2006}), infalling gas is shock-heated. The ensuing hot halo, which constitutes the volume-filling phase of the circumgalactic medium (CGM), suppresses star formation in cluster-resident galaxies. At higher redshifts ($z>1.5$), dense, cold streams can penetrate the hot halo and support star formation \citep{Katz_how_2003, keres_how_2005,mandelker_instability_2016, bennett_resolving_2020}. Therefore, we expect to identify the progenitors of present-day clusters as highly star-forming (up to 50\% of the cosmic star formation rate density) and dense (on scales of 100 pkpc) collections of galaxies \citep{behroozi_average_2013,chiang_galaxy_2017,yajima_forever22_2021,lim_flamingo_2024,morokuma-matsui_forever22_2025, baxter_quantifying_2025}.

These extremely dense regions of the very early Universe are crucial both for early galaxy evolution \citep{laporte_dust_2017,harikane_silverrush_2019,hashimoto_reionization_2023,arribas_ga-nifs_2024,morishita_accelerated_2025, witten_rising_2025,witten_before_2025,li_epochs_2025,Baker2026} and the reionisation process \citep{laporte_lensed_2022,witstok_inside_2024, chen_impact_2025, lu_mapping_2025,Li2026}. The massive haloes within which these galaxy clusters reside represent the most extreme nodes of the cosmic web. They are fed by vast filamentary structures, resulting in protocluster-resident galaxies that often show excessive damping from dense neutral hydrogen \citep[$N_{\rm HI}>10^{22}\ \rm cm^{-2}$;][]{witten_before_2025,Li2026}. 
This gas accretion can drive enhanced and prolonged star formation, which, together with the early formation times of protocluster galaxies, leads to the accelerated evolution of galaxies resident in these dense environments \citep{thomas_epochs_2005}. Protocluster-resident galaxies have increased metallicities, higher dust contents, and older stellar populations relative to field galaxies \citep{laporte_dust_2017,arribas_ga-nifs_2024,witten_rising_2025, li_epochs_2025, helton_jwst_2024}. However, while some protocluster resident galaxies show extreme, evolved properties that are atypical compared to the general high-redshift galaxy population \citep[e.g.][]{roberts-borsani_between_2024}, this is not ubiquitous across high-redshift protoclusters \citep{laporte_lensed_2022,fudamoto_sapphires_2025,Li2026}. The wide variety in the properties of protocluster-resident galaxies demonstrates varying levels of impact from environmental effects. To understand what drives these variations, it is crucial to identify and characterise a larger sample of protoclusters across a diversity of evolutionary phases and redshifts.

In the era of \jwst, with its deep imaging and spectroscopic abilities, discoveries of high-redshift overdense environments have been plentiful. These range from those serendipitously discovered in Near Infrared Camera (NIRCam) imaging or wide-field slitless spectroscopy \citep{fudamoto_sapphires_2025, castellano_early_2023, helton_jwst_2024, helton_identification_2024} to those found in targeted searches around objects that are thought to be ideal tracers of massive DM haloes, including quasars \citep[e.g.][]{kashino_eiger_2023,meyer_jwst_2024,wang_massive_2024, pudoka_large-scale_2024,pudoka_lyman-break_2025,champagne_quasar-anchored_2025}, Lyman-$\alpha$ emitters \citep{witstok_inside_2024}, massive dusty galaxies \citep{arribas_ga-nifs_2024}, and UV-bright galaxies \citep{laporte_lensed_2022, hashimoto_reionization_2023, morishita_early_2023, tacchella_jades_2023,scholtz_gn-z11_2024}. In particular, UV-bright galaxies will become crucial signposts of overdense environments as we move towards wide-area, relatively shallow imaging from instruments such as the {\it Euclid} and {\it Roman} space telescopes, which can detect these rare UV-bright objects. One of the most promising candidate protoclusters is the environment around two UV-bright galaxies in the SMACS J0723.3-7327 (hereafter SMACS0723) field at $z=7.66$ \citep{laporte_lensed_2022}.

In this paper, we utilise new NIRCam imaging and grism observations (discussed in Section~\ref{sec:data}) around this candidate protocluster to build a spectroscopic and photometric sample of protocluster-resident galaxies (Section~\ref{sec:sample}). We then perform spectral energy distribution (SED) fitting on the sample to establish the properties of resident galaxies (Section~\ref{sec:props}). We use the TNG-Cluster and TNG300 simulations to estimate the virial mass of this protocluster and its projected current-day virial mass (Section~\ref{sec:halo_mass}). We establish a connection between the virial masses of high-redshift protoclusters and the impact of environmental effects (Section~\ref{sec:enh_env_eff}), and finally, in Section~\ref{sec:conclusions}, we present our conclusions. 

We corrected all properties for lensing using the magnification factor estimated from \cite{Pascale2022}, unless otherwise stated.

\begin{figure*}%
\centering
\includegraphics[width=0.8\textwidth]{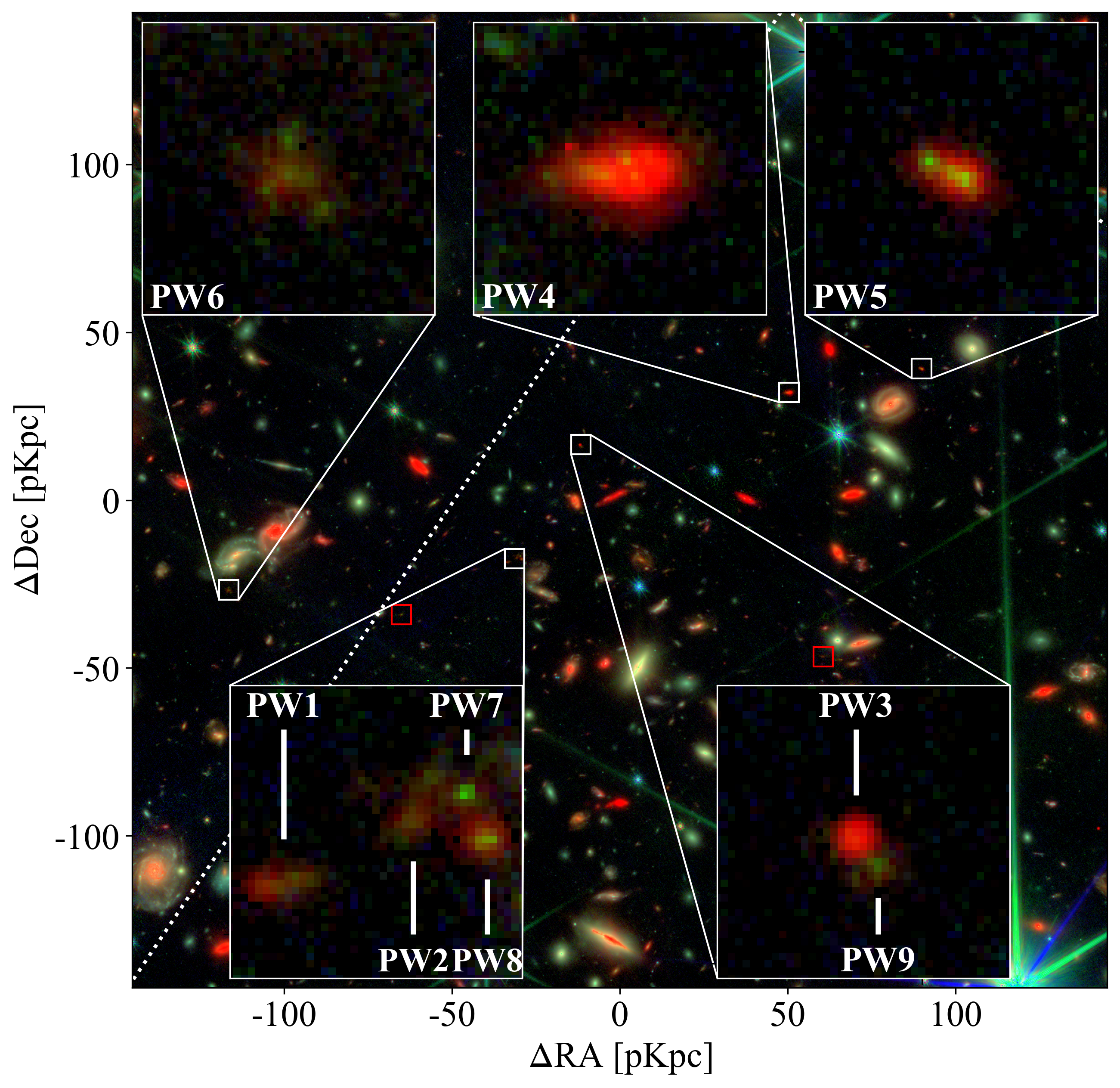}
\caption{Red-green-blue (RGB) image using the F090W, F115W, and F444W filters, showing the protocluster region in the SMACS0723 field. The dashed white line indicates the edge of the ERO imaging, where PW6 is located in a region newly imaged by GO 4043. The inset panels zoom in on regions that contain spectroscopically confirmed $z=7.66$ galaxies. Within the inset panels ($\sim 5\times 5\ {\rm pkpc^2}$), the names correspond to the galaxies indicated by the markers. The red boxes indicate our two lower-confidence photometric candidates. The axes are corrected for lensing using the median magnification, $\mu = 1.43$.}
\label{fig:RGB}
\end{figure*}

 \begin{table*}
    \centering
    \caption{Properties of the \pc-resident galaxy sample.}
    \begin{tabular}{lcccccccc}
        \hline
    ID & RA & Dec. & $z$ & $M_{\rm UV}$ & $\beta$ & $\log(M_{\star}[{\rm M_{\odot}}])$ & $N_{\rm HI}$ & $\mu$ \\
    \hline
    \multicolumn{4}{l}{\bf Spectroscopically confirmed galaxies}&&&\\
    \hline
    PW1 & 110.86717 & -73.43894 &   7.6653    & $-18.6 \pm 0.2$ & $-1.8\pm 0.3$ & $8.6^{+0.2}_{-0.2}$ & $23.3 \pm 0.2$ & 1.47 \\
    PW3 & 110.86132 & -73.43627 & 7.6698 & $-17.9 \pm 0.2$ & $-1.4\pm 0.2$ & NA & NA & 1.40 \\
    PW4 & 110.84460 & -73.43508 & 7.6661 & $-20.36\pm0.04$ & $-1.4 \pm 0.1$ &$9.2^{+0.2}_{-0.2}$ & $<21.5$&1.38 \\
    PW5 & 110.83397 & -73.43454 & 7.6624 & $-19.95\pm 0.04$ & $-2.1\pm 0.1$ & $8.1^{+0.2}_{-0.1}$ &$<21.5$& 1.39 \\
    PW6 & 110.88945 & -73.43958 & 7.6625 & $-19.72\pm 0.03$ & NA & $8.7^{+0.2}_{-0.2}$ & NA &1.43 \\
    PW8 & 110.86599 & -73.43887 & 7.6616 & $-18.79 \pm 0.09$ & $-2.1 \pm 0.2$ & $7.9^{+0.2}_{-0.3}$& $22.7 \pm 0.2$& 1.47 \\
    \hline
    \multicolumn{5}{l}{\bf Close companion photometric candidates}&&\\
    \hline
    PW2 & 110.86638 & -73.43884 & $7.7^{+0.5}_{-0.6}$ & $-17.8\pm0.2$ & $-1.8\pm0.3$ & $8.4^{+0.3}_{-0.4}$ & $<21.5$&1.47 \\
    PW7 & 110.86610 & -73.43880 & $7.4^{+0.5}_{-0.3}$ & $-18.5\pm 0.1$ & $-2.1\pm 0.2$ & $8.3^{+0.2}_{-0.3}$  &$<21.5$& 1.47 \\
    PW9 & 110.86118 & -73.4363 &$7.3^{+0.6}_{-0.2}$ &$-17.8\pm 0.2$ & $-1.8 \pm 0.4$&$8.1^{+0.3}_{-0.6}$&$<21.5$&1.40\\
    \hline 
    \multicolumn{5}{l}{\bf Tentative photometric candidates}&&\\
    \hline
    PW12 & 110.84191 & -73.44109 & $7.8^{+0.4}_{-0.7}$ &$-18.6 \pm 0.1$& $-1.9 \pm 0.2$& $8.1^{+0.3}_{-0.2}$&$<21.5$&1.62\\
    PW13 & 110.87559 &-73.44013 & $7.2^{+0.5}_{-0.3}$&$-18.8\pm 0.2$&$-2.2\pm0.3$&$8.6^{+0.1}_{-0.2}$&$<21.5$&1.49\\
    \hline
    \end{tabular}
    \\
    Notes: The table is divided into a spectroscopic sample, a close companion photometric candidate sample, and a sample of photometric candidates that require future confirmation. The magnification factors are from \cite{Pascale2022}.
    \label{tab:gal_props}
\end{table*}

\section{Data}
\label{sec:data}
The SMACS0723 field was the subject of the first \jwst\ deep field as part of the Early Release Observations \citep[ERO; ID: 2736,][]{pontoppidan_jwst_2022}, including Near Infrared Spectrograph (NIRSpec), Near Infrared Camera (NIRCam), and Near Infrared Imager and Slitless Spectrograph (NIRISS) observations. Observations with the NIRCam instrument using the F090W, F150W, F200W, F277W, F356W, and F444W filters revealed a number of high-redshift ($z>7$) photometric candidate galaxies \citep[][we refer the reader to these works for details on the imaging depths of the ERO data]{morishita_physical_2023,donnan_evolution_2023, atek_revealing_2023,adams_discovery_2023,harikane_comprehensive_2023}, while NIRSpec observations with the disperser-filter combinations G235M/F170LP and G395M/F290LP spectroscopically confirm a small sample of such objects \citep{curti_chemical_2023, brinchmann_high-z_2023, tacchella_stellar_2022, carnall_first_2023,trump_physical_2023}. The original ERO spectroscopy and imaging of two $z=7.66$ galaxies suggest they have typical metallicities given their stellar masses \citep{curti_chemical_2023}, are young with rising star formation histories (SFHs) and clumpy morphologies \citep{tacchella_jwst_2023}, and may host active galactic nuclei \citep[AGNs; ][]{brinchmann_high-z_2023}, which drive subsequently detected outflows \citep{Ivey2026}. These two UV-bright ($M_{\rm UV}\sim -20$) galaxies are separated by less than 50 pkpc (see PW4 and PW5 in Figure~\ref{fig:RGB}), leading to the hypothesis that they likely reside within a massive DM halo. Further supported by an overdensity of photometrically identified galaxies, this region of the SMACS0723 field appears to be a protocluster core candidate at $z=7.66$ \citep{laporte_lensed_2022}, hereafter \pc.

The Build-Up of Large-Scale Structure in the Early Universe Cycle 2 GO programme (ID: 4043, PI: Witten) aimed to spectroscopically confirm the photometric protocluster candidate, \pc. The programme used simultaneous NIRCam wide-field slitless spectroscopy (WFSS) with the F444W filter, together with complementary short-wavelength imaging in the F090W ($5\sigma$ limiting magnitude of 29.1 AB) and F115W filters ($5\sigma$ limiting magnitude of 29.5 AB). These observations were conducted with both the R and C grism directions to break the spatial-spectral degeneracy of WFSS and to provide a second dispersal direction for when the spectrum is heavily contaminated \citep[e.g.][]{naidu_all_2024}. This contamination is especially common when the dispersion passes through the foreground cluster region. The programme primarily aimed to detect \OIII-emitters at $z=7.66$, within \pc\ and improve the photometric redshifts of candidate galaxies. Covering a single NIRCam pointing, the 15890 s integration time in each grism direction makes this programme one of the deepest \jwst\ grism observations ever performed.

The simultaneous short-wavelength NIRCam imaging in F115W covers a wavelength range ($1.3\lesssim\lambda\lesssim1.7$) that was previously poorly constrained by existing, relatively shallow, {\it Hubble} Space Telescope imaging and was not included in the ERO filter configuration. This wavelength range is crucial for estimating the redshift of $7<z<9$ photometric candidate galaxies, and as such there is a significant overlap between the $z\sim 9 $ photometric candidates in the literature \citep[e.g.][]{donnan_evolution_2023} and our now spectroscopically confirmed $z=7.66$ galaxies.

Throughout this paper we use the NIRCam imaging data reductions from the Dawn \jwst\ Archive (DJA)\footnote{\url{https://dawn-cph.github.io/dja/}}. The DJA retrieves level-two calibrated NIRCam exposures and processes them using the \texttt{grizli} software package \citep{brammer_grizli_2019} to derive combined, drizzled images that are aligned to Gaia astrometry (for further information, see \citealt{valentino_atlas_2023}). We processed the NIRCam grism data in the same way using the \texttt{grizli}\footnote{\url{https://github.com/gbrammer/grizli}} code, with standard calibrations but with grism configuration files (V9) provided by Nor Pirzkal\footnote{\url{https://github.com/npirzkal/GRISMCONF}}. We then followed the same procedure as described in \citet{oesch_jwst_2023} and \citet{meyer_jwst_2024}, using standard Calibration References Data System (CRDS) grism dispersion files, the \texttt{grizli} sensitivity function, and median filtering to extract continuum-subtracted grism spectra.

\section{Protocluster-resident sample}
\label{sec:sample}
\subsection{Spectroscopic sample}
Given the newly available NIRCam WFSS observations, we primarily targetted emission-line galaxies. We began with a basic Lyman-$\alpha$ break selection criterion for $z\sim 7$ galaxies of F090W - F115W $> 1.5$ (following the dropout criterion of \citealt{Weibel2025}). We then required both a non-detection at $3\sigma$ depth in F090W and a detection at greater than $3\sigma$ in F444W. We chose this liberal $z\sim 7$ selection criterion because parts of the SMACS0723 field that are newly imaged by this programme are covered only by the F090W, F115W, and F444W filters, making a selection based on these filters necessary. 

\begin{figure}
    \centering
    \includegraphics[width=1\linewidth]{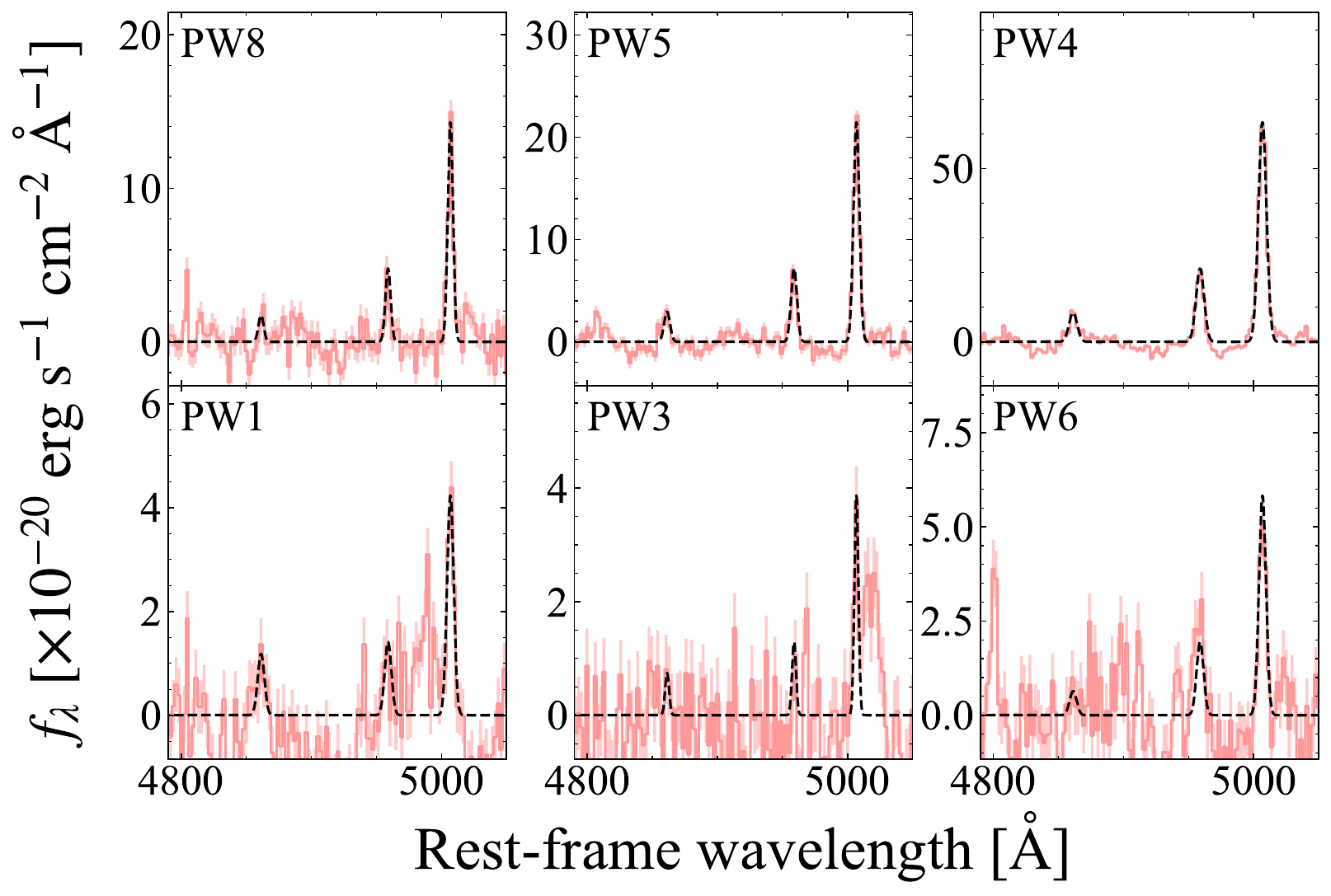}
    \caption{Rest-frame spectra of \OIII-emitters identified in the SMACS0723 field at $z=7.66$. Each panel shows a different galaxy, identified by the text in the top left of each panel. The shaded red region around the red solid line shows the error of each spectrum. The dashed line shows the best-fit model spectrum of the \Hb, [\ion{O}{III}] $\lambda$4960\AA\ and [\ion{O}{III}] $\lambda$5008\AA\ emission lines. Despite the relative weakness of the emission lines in PW3 and PW1, their detection in the 2D spectra in both R and C directions, together with their consistent photometric redshifts, indicates that they are emission lines at $z=7.66$.}
    \label{fig:ELs}
\end{figure}

This selection resulted in a sample of $\sim 2600$ objects that required visual inspection. We used the \texttt{specvizitor} graphical user interface\footnote{\url{https://github.com/ivkram/specvizitor}} to inspect potential \OIII\ emitting galaxies at $z\sim7.66$. We searched both the 1D and 2D grism spectra for emission lines in the wavelength range $4.2 \, \mu{\rm m}<\lambda_{\rm obs}<4.4 \, \mu{\rm m}$ and identified a total of six spectroscopically confirmed galaxies at $z=7.66$. We fit their \OIII\ and \Hb emission lines with Gaussian profiles and Monte Carlo error propagation, and we obtained the fluxes reported in Table~\ref{tab:fluxes} and the redshifts reported in Table~\ref{tab:gal_props}. We show the combined spectrum from the R and C grism directions, as well as our best-fit emission line model, in Figure~\ref{fig:ELs} for each galaxy. We note that while the emission line detections for PW3 and PW1 appear to be tentative in the combined 1D spectra, our emission line modelling identifies them at greater than or equal to $3\sigma$. In addition, they appear present in both the R and C 2D spectra and have SEDs that are consistent with a Lyman-$\alpha$ break at $z\sim 8$. Therefore, we consider these galaxies as spectroscopically confirmed. 

\subsection{Photometric sample}

We calculated the photometry of each galaxy using the routine described in \cite{witten_before_2025}. This routine uses the \texttt{Photutils} package \citep{larry_bradley_astropyphotutils_2025} with custom apertures and local background subtraction to avoid contamination from the intracluster-light of the foreground cluster (we report the photometry of each galaxy in Table~\ref{tab:phot}).

We also searched for potential photometric candidates in the immediate environment of our spectroscopic sample that may not be deblended by automated photometric extraction pipelines. We therefore extracted the photometry of nearby sources with our \texttt{Photutils} routine, discussed above, and identify three candidates, PW2, PW7 and PW9, that are resident within 2 pkpc of a spectroscopically confirmed galaxy and have an SED consistent with a $z\sim8$ galaxy. We fit the photometry of these galaxies (see Table~\ref{tab:phot}) with the photometric redshift tool \texttt{Eazy} \citep{brammer_eazy_2008}, using the \texttt{blue\_sfhz} template set. We report these photometric redshifts in Table~\ref{tab:gal_props} and are consistent with $z=7.66$, with a probability of less than $0.1\%$ of having a redshift outside $7<z_{\rm phot}<8.5$. Given this and their very close proximity to nearby spectroscopically confirmed galaxies, we consider it very likely that they are at coincident redshifts, and thus these become part of our higher-confidence, close-companion photometric sample. Finally, we used the automated \texttt{Eazy} photometric redshift catalogue released on the DJA\footnote{\url{https://dawn-cph.github.io/dja/imaging/v7/}}, to search for further photometric candidates across the full SMACS0723 field. We identify two robust high-redshift candidates that have photometric redshifts consistent with $z=7.66$. For consistency, we re-extracted their photometry using the same \texttt{Photutils} routine and refitted their photometric redshifts using the same \texttt{Eazy} templates as used for our `close companion photometric sample'. Tables~\ref{tab:phot} and ~\ref{tab:gal_props} show the resultant photometry and photometric redshifts, respectively. The lack of medium-band observations in the SMACS0723 field makes photometric redshift estimates relatively uncertain \citep[e.g.][]{arrabal_haro_confirmation_2023}. As such, we consider these as objects of lower-confidence that require follow-up confirmation. Figure~\ref{fig:RGB} shows the positions of our sample in the RGB image, all of which lie within a $\sim 100 \times 200\ {\rm pkpc}^2$ region. 

Finally, we note that the photometry of PW3 satisfies the photometric diagnostics of the so-called little red dots (LRDs; \citep{matthee_little_2024}), which is discussed in detail in Appendix~\ref{app:LRD}. The strong Balmer breaks and red rest-optical slopes of LRDs are likely driven by dense gas around a central black hole \citep{inayoshi_extremely_2025, ji_blackthunder_2025, deugenio_blackthunder_2025,naidu_black_2025}. We therefore did not perform SED fitting with stellar models for this source, as detailed in the following section, and do not analyse it in the same way as for the other galaxies in our sample.

\subsection{Confirmation of an overdensity}

Most protoclusters are identified through an excess of UV-bright galaxies within small volumes of the early Universe. These are often characterised by the number of observed galaxies relative to the expected number from the integral of the UV luminosity function ($\delta + 1 = N_{\rm obs}/N_{\rm exp}$). To measure the UV luminosity and the UV slope of our sample, we fitted the F150W, F200W, and F277W filters with a power law ($f_{\lambda} \propto \lambda ^{\beta}$), with Monte Carlo uncertainty propagation and evaluated the model at $\lambda=0.15 \mu$m. Table~\ref{tab:gal_props} lists the inferred UV slopes and luminosities.

To estimate the UV overdensity, we used a volume encompassing our entire sample, defined by a radius of $R = 0.1$ pMpc and a redshift interval of $7.65<z<7.67$ ($\sim 0.7$ pMpc), resulting in a volume of $\sim 19\ {\rm cMpc}^3$. We integrated the UVLF to the $5 \sigma$ limiting magnitude ($m_{\rm F150W}\sim 29.7\ {\rm AB}$) within this volume, yielding $\delta \sim 200$. 

\begin{figure}%
\centering
\includegraphics[width=0.45\textwidth]{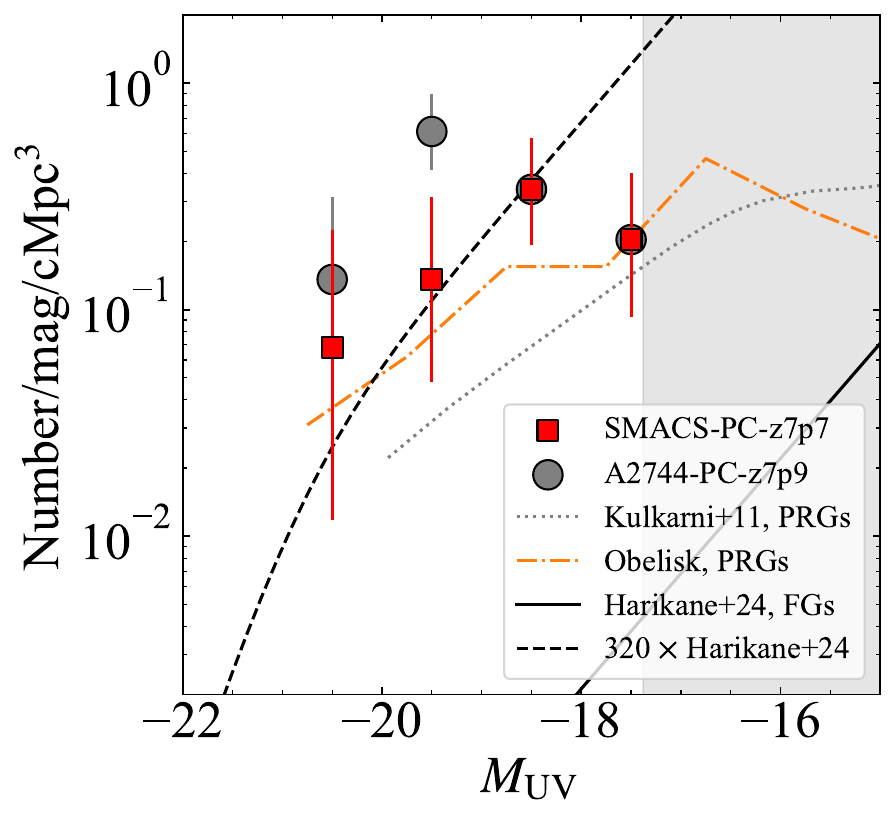}
\caption{UV luminosity function of galaxies within \pc\ (red squares) compared with the UVLF of another $z\sim8$ protocluster, A2744-PC-z7p9 \citep[][grey circular points]{witten_before_2025}, both calculated within the same central volume. The UVLFs of other overdense environments from simulations are shown \citep[][dash-dotted orange and dotted grey  lines, respectively]{trebitsch_obelisk_2021,kulkarni_reionization_2011} and have a similar shallow gradient to that seen in \pc. The shaded grey region represents the $5\sigma$ depth in the UV-continuum, beyond which we expect the UVLF to become highly incomplete. Relative to the nominal UVLF of $z\sim 8$ galaxies from \cite{harikane_pure_2024} (solid black line), the UVLF of \pc\ appears to be $\sim 300$ times higher (dashed black line).}
\label{fig:LF}
\end{figure}

Figure~\ref{fig:LF} shows the UVLF of \pc. We compared it to the nominal UVLF \citep{harikane_pure_2024} and find it to be $\sim 300$ times higher than the nominal UVLF. We note here that while we expect the UVLF to become incomplete near the $5\sigma$ imaging depth, correcting for completeness would only further enhance the offset from the nominal UVLF. We additionally compared the UVLF of \pc\ with that of A2744-PC-z7p9, measured over the same central volume of $\sim 19\ {\rm cMpc}^3$. The normalisation of the local UVLF of A2744-PC-z7p9 is around a factor of six greater than that of \pc\ in the regime where the photometric selection has a high completeness, indicating that \pc\ is a slightly less extreme environment than A2744-PC-z7p9.

We also compared the UVLF of \pc\ to simulations of overdense environments from \cite{kulkarni_reionization_2011} and \cite{trebitsch_obelisk_2021}\footnote{The UVLF of Obelisk at $z=7.9$, shown in Figure~\ref{fig:LF}, is taken from \cite{witten_before_2025}, which focuses on the central $R=2$~cMpc of the protocluster and applies a dust prescription with a Small Magellanic Cloud-like extinction law, described in Trebitsch et al. in prep. and \citet{volonteri_exploring_2025}.}. Both simulations show a shallower evolution of the UVLF relative to the nominal UVLF. They also show an excess of bright objects, which are known to be biased toward overdense environments \citep{jespersen_significance_2025}, and a deficit of fainter objects. This deficit is expected due to enhanced Lyman-Werner production in these environments, which suppresses star formation in low-mass DM haloes. The SMACS0723 UVLF is consistent with the shape of both the nominal UVLF (scaled by a factor of 300) and the shallower UVLF from simulations of overdense environments; however, we do note a slight excess of UV-bright galaxies, in line with the latter. 

Given the small volumes being considered, it is important to quantify the effects of cosmic variance. We used {\it QuickCV} \citep{newman_measuring_2002} to estimate the DM variance of our volume ($24\%$), and assuming a galaxy bias of $b_g = 8$ from \cite{shuntov_constraints_2025}, we find that at $1\sigma$ cosmic variance can account for a tripling in the expected number of galaxies within this small volume. As such, our observed overdensity, $\delta=200$, is an $\sim 60\sigma$ outlier relative to cosmic variance. If we instead consider a significantly larger volume than that occupied by the \pc\ galaxies, corresponding to the Lagrangian radius from \cite{chiang_galaxy_2017} ($R_{L}=10\ {\rm cMpc}$), the overdensity falls to $\delta = 3$, which is a $2.6 \sigma$ outlier relative to cosmic variance. We note that the overdensity parameter of A2744-PC-z7p9, when calculated over the same large volume, falls to a similarly low value of $\delta \sim 8$. This large volume is far in excess of the expected sizes of protocluster cores at these redshifts \citep{chiang_galaxy_2017}. Therefore, regardless of the volume selected, these regions appear to be significantly overdense relative to the nominal UVLF.

While establishing the UV luminosity of our sample, we also obtained the UV slope of each galaxy (except for PW6, which resides in the newly imaged region of SMACS0723 and thus only has F090W, F115W, and F444W imaging available). This reveals a diversity in the UV slopes of our sample, ranging from $-2.2<\beta<-1.4$. The majority of \pc\ galaxies have UV slopes that are largely consistent with the average UV slope measured from photometry at this redshift \citep[$\beta\sim -2.2$ at $z\sim 7.6$;][]{austin_epochs_2024}. These are consistent with the UV slopes measured for non-core resident galaxies in A2744-PC-z7p9. However, two galaxies in our sample, PW3 and PW4, show distinctly redder UV slopes, $\beta = -1.4$. As discussed earlier, PW3 is likely an LRD and hence we do not interpret its UV slope. PW4 is the most UV-luminous galaxy in our sample. While a mild increase in the UV slope is expected with $M_{\rm UV}$ \citep[e.g.][]{saxena_hitting_2024}, this object is redder than expected given its UV luminosity. Based on its luminosity and red UV slope, we hypothesise that PW4 likely represents the central galaxy resident within the core of this protocluster, akin to the massive core galaxies with red UV slopes reported for A2744-PC-z7p9 in \cite{witten_before_2025}.

\section{Galaxy properties}
\label{sec:props}

We used the SED-fitting code \texttt{Prospector} \citep{johnson_stellar_2021} to simultaneously fit the photometric data and the \Hb and [\ion{O}{III}] emission-line fluxes. For our photometric sample, we used the upper limits on the \Hb and [\ion{O}{III}] fluxes and fixed the redshift to that of the protocluster, $z=7.66$. We used the same 16-parameter model as \cite{witten_rising_2025} \citep[based on][]{tacchella_jwst_2023}, which most notably includes a non-parametric SFH with a continuity prior \citep{leja_how_2019}. The SFH accounts for ten of these parameters, including the star formation rate (SFR) ratio between bins and the total stellar mass of the source. Additionally, this model includes a flexible two-component dust attenuation model \citep{charlot_simple_2000}, with the following components: attenuation from a birth cloud, applied to young stars, and attenuation from a diffuse component with a flexible power-law slope. The remaining parameters are the stellar metallicity, the gas phase metallicity, and the ionisation parameter.

Following \cite{witten_before_2025}, we avoided fitting the NIRCam filter covering the Lyman-$\alpha$ break. This is due to the potential for excessive damping around the Lyman-$\alpha$ break caused by significant neutral hydrogen column densities \citep[e.g.][]{heintz_strong_2024}. This damping is not modelled by \texttt{Prospector} and is especially prominent in massive DM haloes and hence in overdense environments \citep{witten_rising_2025,gelli_neutral_2025,Heintz2026,Li2026}. Therefore, we did not include the F115W filter in our \texttt{Prospector} SED-fitting. 

Table~\ref{tab:gal_props} lists the stellar masses inferred from this SED fitting and Figure~\ref{fig:SEDs} shows the best-fit SED and SFH. We omit the SED-fitting results for PW3 and discuss them further in Section~\ref{sec:LRD}. Moreover, in some of the following analysis we omit the galaxy PW6 as it lies outside of the original ERO photometry. PW6 is therefore only detected in two \jwst\ filters, meaning the inferred galaxy properties should be treated with caution.

\subsection{Neutral hydrogen column densities}

\begin{figure}%
\centering
\includegraphics[width=0.5\textwidth]{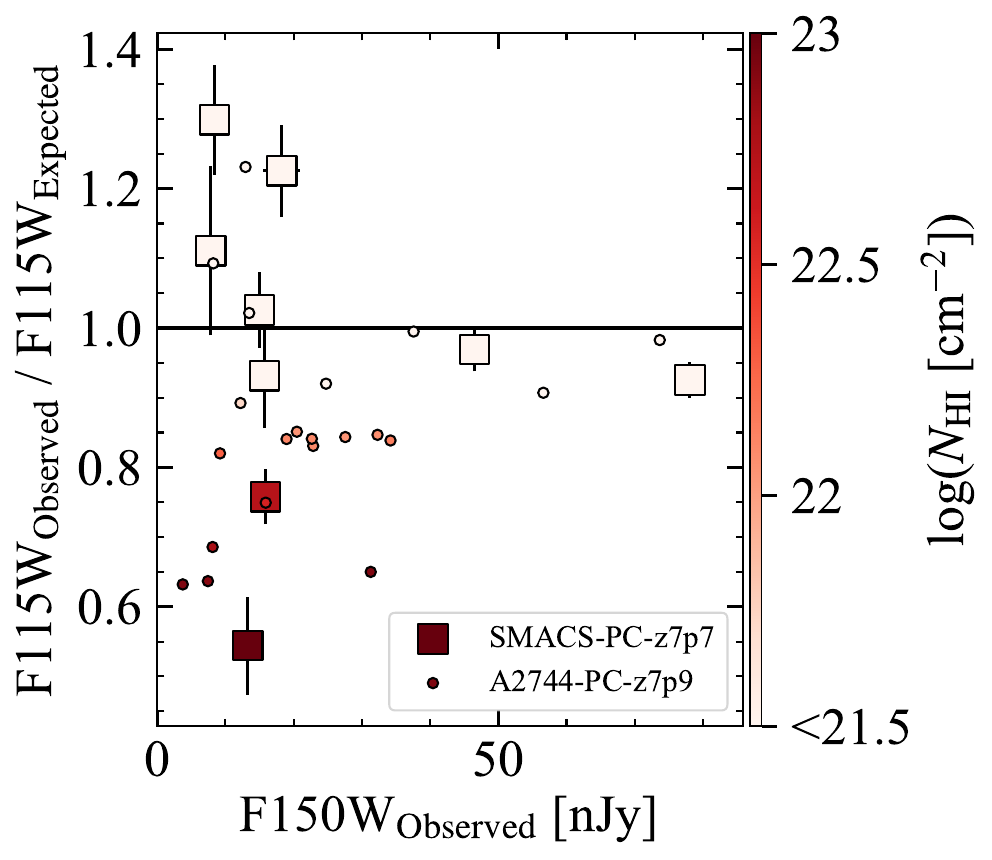}
\caption{Ratio of observed to expected flux in the F115W filter. Increasing neutral hydrogen column density, shown by the colour bar, produces excessive damping of the Lyman-$\alpha$ break, thereby suppressing the observed flux in the F115W filter. The photometry of \pc-resident galaxies is shown by squares, while A2744-PC-z7p9-resident galaxies are shown as circles \citep{witten_before_2025}. While two galaxies in \pc\ show extreme column densities, this represents a much smaller fraction of the sample than those with extreme column densities in A2744-PC-z7p9.}
\label{fig:NHi}
\end{figure}

As discussed above, we intentionally omitted the F115W filter from our SED fitting given that we anticipated potential damping from high neutral hydrogen column densities, which is not modelled by \texttt{Prospector}. We then used our best-fitting \texttt{Prospector} model to estimate the required damping, and hence the neutral hydrogen column density, to match the observed F115W flux. We first applied an intergalactic medium (IGM) attenuation curve to the model, assuming a neutral fraction of one and no ionised bubble following \cite{mason_measuring_2020}, which maximises the impact of the IGM and hence minimises the impact of the damped \Lya\ (DLA) (thus our $N_{\rm HI}$ estimate is technically a lower bound). We then assumed a Voigt-Hjerting profile for the DLA absorption, which is valid for high $N_{\rm HI}$, using Equation nine of \cite{smith_lyman_2015}. Finally, we iterated over a grid of $N_{\rm HI}$ to find the required damping of the model spectrum to match the observed F115W flux. Figure~\ref{fig:NHi} shows the results and are listed in Table~\ref{tab:gal_props}. 

While PW1 and PW8 have inferred $N_{\rm HI}>10^{21.5}\ {\rm cm^{-2}}$, this amounts to only $\sim 20 \%$ of the sample, compared to $\sim 60 \%$ of galaxies in A2744-PC-z7p9. Notably, the galaxies impacted by the highest neutral hydrogen column density are located within the most clustered region of \pc, which includes PW1, PW2, PW7, and PW8. A similar trend is also seen in A2744-PC-z7p9 \citep{witten_before_2025}. Two of the \pc-resident galaxies show excess emission relative to the model spectrum in the F115W filter, beyond the observational uncertainties. This excess can be due either to uncertainties in our \texttt{Prospector} models or to \Lya emission contaminating the F115W filter. 

\subsection{Star formation histories}

\begin{figure}%
\centering
\includegraphics[width=0.5\textwidth]{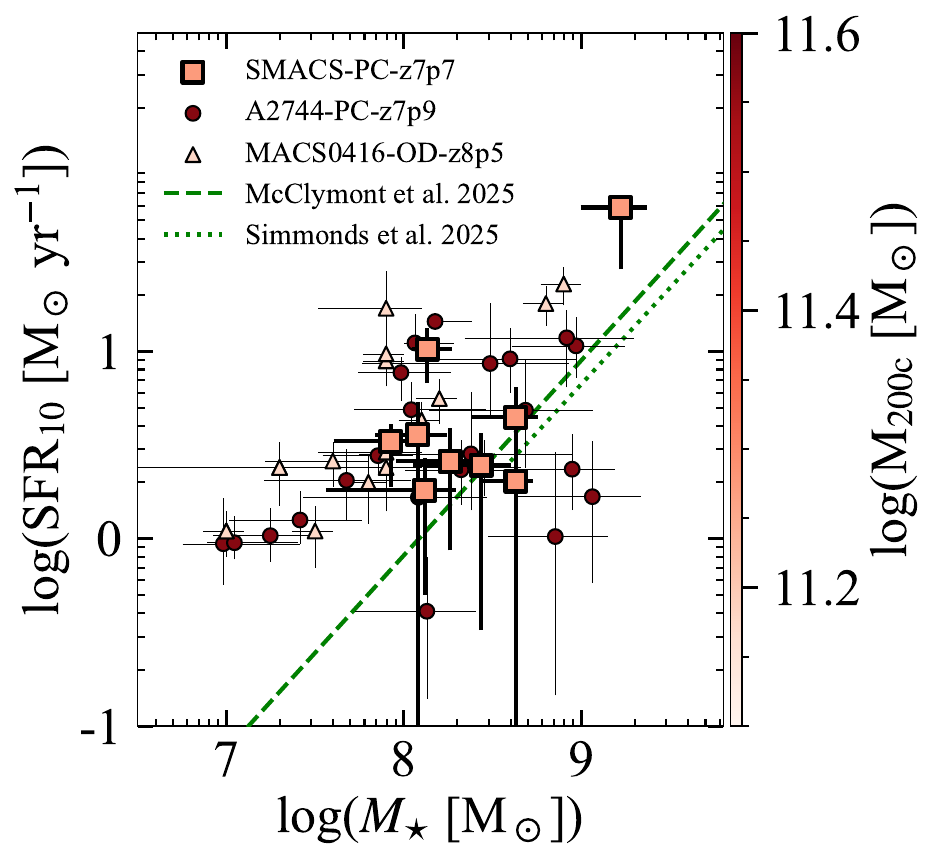}
\caption{Star-forming main sequence (SFMS) of galaxies in \pc\ (squares), A2744-PC-z7p9 \citep[circles;][]{witten_before_2025} and MACS0416-OD-z8p5 \citep[triangles;][]{fudamoto_sapphires_2025}, compared to the SFMS of field galaxies measured from observations \citep[][dotted green line]{simmonds_bursting_2025} and simulations \citep[][dashed green line]{mcclymont_thesan-zoom_2025}. The SFMS is coloured based on its virial mass, estimated using the method discussed in Section~\ref{sec:halo_mass}. While the most massive galaxies in A2744-PC-z7p9 are below the SFMS, all galaxies in \pc\ and MACS0416-OD-z8p5 fall on or above the SFMS of field galaxies.}
\label{fig:SFMS}
\end{figure}

By jointly fitting photometry and our \OIII\ and \Hb emission line constraints, we can break the degeneracy between the Balmer break and strong emission lines scenarios \citep[e.g.][]{roberts-borsani_interpreting_2020}. With the rest-optical continuum measurement, we can place improved constraints on the SFHs of resident galaxies. While some of our galaxies (PW2, PW7, and PW13) show evidence of a mild Balmer break, their SFHs remain consistent with a recent, ongoing burst of star formation (Figure~\ref{fig:SEDs}), as previously suggested for PW4 and PW5 \citep{tacchella_jwst_2023}. These objects exclusively show ${\rm SFR_{10}/SFR_{100}>1}$ (where ${\rm SFR}_{10}$ and ${\rm SFR}_{100}$ refer to the average SFR over the last 10 Myr and 100 Myr, respectively), i.e. rising SFHs; this contrasts with the wide range of SFHs in A2744-PC-z7p9 \citep{witten_before_2025}. 

Using these inferred SFHs, we placed \pc-resident galaxies on the star-forming main sequence (SFMS) in Figure~\ref{fig:SFMS}. We compared these to protocluster-resident galaxies from A2744-PC-z7p9 \citep{witten_before_2025} and MACS0416-OD-z8p5 \citep{fudamoto_sapphires_2025}, and to field galaxies from observations \citep{simmonds_bursting_2025} and simulations \citep{mcclymont_thesan-zoom_2025}. Notably, the entire \pc\ galaxy sample is consistent with, or above, the SFMS of field galaxies. While observations are biased toward detecting highly star-forming low-mass galaxies, they are highly complete at the high-mass end ($M_{\star}\gtrsim10^9\ {\rm M_{\odot}}$) thanks to deep NIRCam imaging \citep{simmonds_bursting_2025}, as is the case for both SMACS0723 and Abell 2744 \citep[both with similar F115W imaging depths;][]{bezanson_jwst_2024}. Therefore, the lack of galaxies below the SFMS at high masses is noteworthy. This contrasts with galaxies in A2744-PC-z7p9, which show large scatter around the SFMS and, at the highest masses, fall below it. This suggests that instead of suppressing star formation, as in some galaxies resident in A2744-PC-z7p9, residency within \pc\ actually enhances SFRs. This likely results from significant gas accretion that is expected in these environments. This interpretation is supported by the extreme neutral hydrogen column densities detected in PW1 and PW8. Simultaneously, there must not be an onset of significant environmental effects that suppress star formation, as reported for A2744-PC-z7p9 \citep{witten_before_2025}. 

\begin{figure}%
\centering
\includegraphics[width=0.5\textwidth]{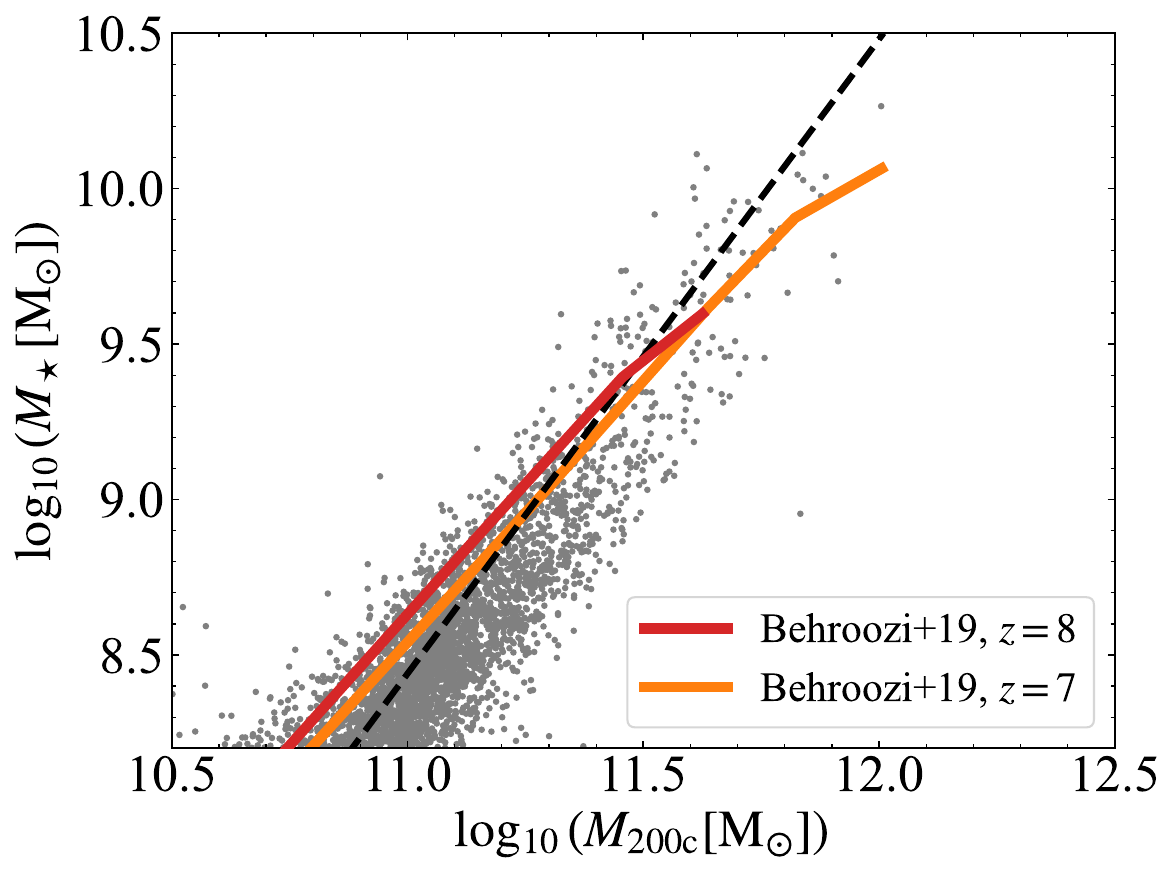}
\caption{Stellar mass within the virial radius as a function of the halo mass for haloes in the TNG300 and TNG-Cluster simulations. Haloes at $z=7.6$ are shown as grey points. We fit these data points with a linear fit, shown by a dashed black line. We also show the stellar mass-halo mass (SMHM) relation from \cite{behroozi_universemachine_2019} at $z=7$ (solid orange line) and $z=8$ (solid red line). Our adapted SMHM relation provides a diagnostic for the virial mass based on the stellar mass within the virial radius.}
\label{fig:Halo_stellar_mass_relation}
\end{figure}

\begin{figure*}%
\centering
\includegraphics[width=0.8\textwidth]{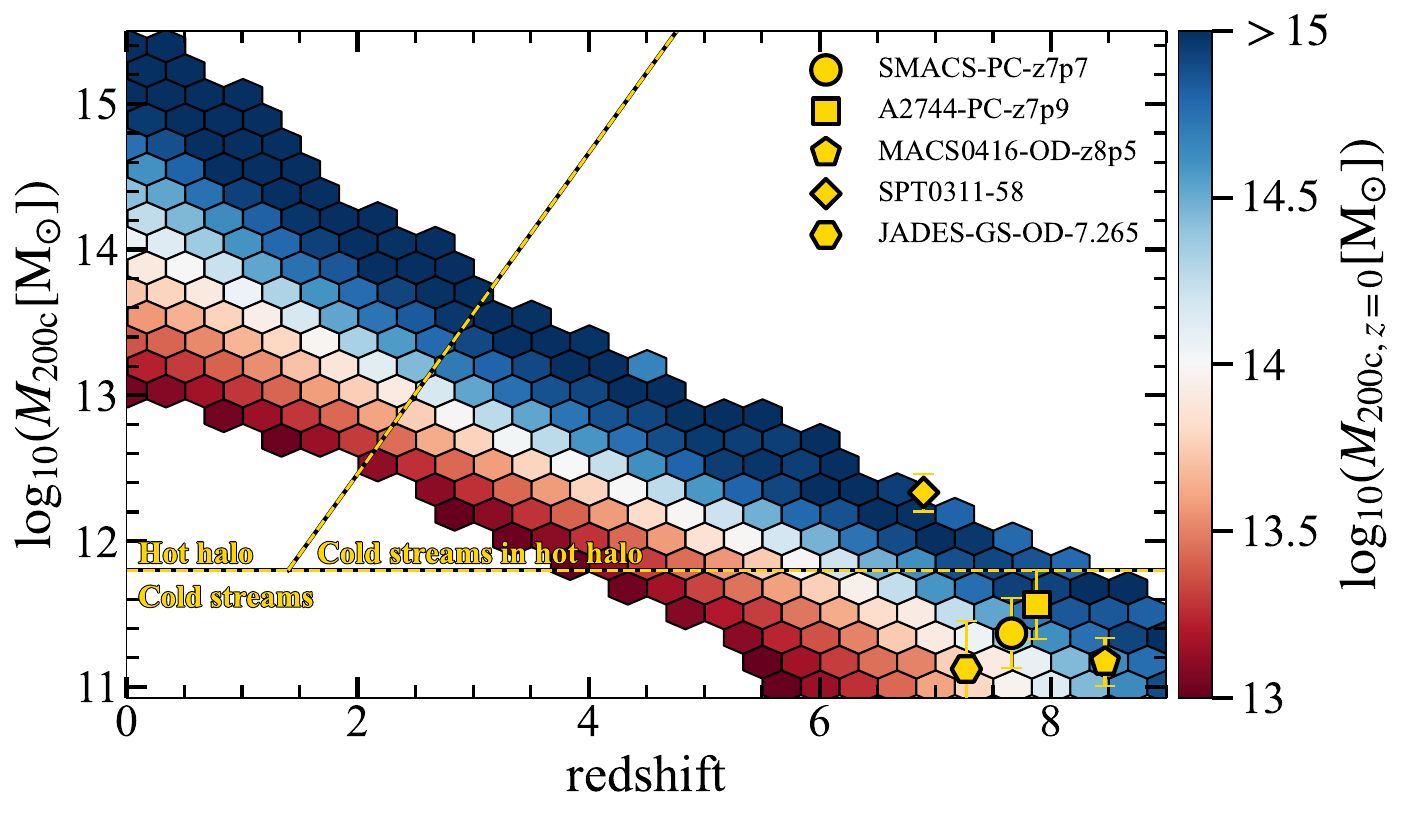}
\caption{Present-day virial masses of haloes in the TNG-Cluster and TNG300 simulations from $0< z < 9$. The colour of the hexagonal bins indicates the median present-day virial mass of the haloes within each bin, but only when the median is $\log_{10}(M_{200{\rm c},\ z=0}\ [{\rm M_{\odot}}]) > 13$. We interpolate to fill voids in the simulations' redshift grid. The different phases of CGM gas are indicated by dashed gold lines from \cite{dekel_galaxy_2006}. The gold data points indicate the estimated virial masses for $z\gtrsim7$ protoclusters with published stellar masses of resident galaxies (circle: SMACS-PC-z7p7, this work; square: A2744-PC-z7p9, \citealt{witten_before_2025}; pentagon: MACS0416-OD-z8p5, \citealt{fudamoto_sapphires_2025}; diamond: SPT0311-58, \citealt{arribas_ga-nifs_2024}; and hexagon: JADES-GS-OD-7.265, \citealt{helton_identification_2024},\citealt{Li2026}).}
\label{fig:Halo_mass_evolution}
\end{figure*}

\section{Halo masses}
\label{sec:halo_mass}
Methods for estimating DM halo masses for protocluster environments at high redshift are notoriously inhomogeneous. Protoclusters are often characterised by dense cores of massive galaxies, so the assumption of a single central massive galaxy is inappropriate. We therefore used the TNG300 \citep{pillepich_first_2018,nelson_first_2018,naiman_first_2018,marinacci_first_2018,springel_first_2018} and TNG-Cluster \citep{nelson_introducing_2024} simulations to study the stellar mass within the virial radius as a diagnostic of halo mass. We calculated this by summing the stellar masses of subhaloes that fall within $R<R_{200\rm c}$ (the virial radius) of the most massive galaxy in the halo. Figure~\ref{fig:Halo_stellar_mass_relation} shows the stellar mass within the virial radius as a function of the virial mass. We fitted the haloes at $z\sim 7.6$ with stellar masses greater than $M_{\rm \star}>10^{8.2}\ {\rm M_{\odot}}$ with a linear fit (${\rm log_{10}}M_{200 \rm c} = a\times {\rm log_{10}}M_{\rm \star} +b$; $a=0.49$, $b=6.86$) and find a scatter of 0.13 dex around the best-fit. We find a mild discrepancy between the \cite{behroozi_universemachine_2019} stellar mass-halo mass (SMHM) relation, which is calculated using only the stellar mass of the central galaxy, in contrast to our relation, which uses the total stellar mass within the virial radius. This offset is also impacted by the underestimation of stellar mass driven by the lower resolution of simulations such as TNG300 and TNG-Cluster \citep[e.g.][]{pillepich_first_2018, vogelsberger_high-redshift_2020}. This can result in offsets of up to 0.3 dex at $z\sim 10$ \citep{kannan_millenniumtng_2023}.

The virial radii of the most massive haloes in TNG300 and TNG-Cluster at $z\sim8$ are $R_{200 \rm c}\lesssim 30$ pkpc; therefore, we used the total stellar mass within the same limiting virial radius centred on the most massive galaxy. In \pc, PW4 represents the most massive galaxy (${\rm log_{10}}(M_{\rm \star} [{\rm M_{\odot}}]) = 9.2^{+0.2}_{-0.2}$), and while an underlying galaxy population below the depth of current imaging is likely, it probably does not contribute significantly to the stellar mass within the virial radius. Using our best-fit linear model, we estimate the virial mass of \pc\ to be ${\rm log_{10}}(M_{200 \rm c} [{\rm M_{\odot}}]) = 11.4^{+0.2}_{-0.2}$, with an uncertainty given by summing the stellar-mass uncertainty and scatter in the stellar-to-halo mass relation in quadrature. We repeated this process for a number of well-known protocluster candidates at $z\gtrsim7$, using reported stellar masses from \cite{fudamoto_sapphires_2025, witten_before_2025,arribas_ga-nifs_2024,Li2026}. We show their inferred virial masses in Figure~\ref{fig:Halo_mass_evolution}. 

To understand how these observed high-redshift massive haloes evolve as a function of time, we tracked the $z=0$ descendants of all haloes with masses of $M_{\rm 200c}>10^{11}\ {\rm M_{\odot}}$ in TNG300 and TNG-Cluster at every snapshot from $0<z<9$. When multiple haloes result in the same descendant at $z=0$, we only included the most massive high-redshift halo in this analysis. We only considered the most massive progenitor haloes at high redshift, as these are comparable to the halo masses of observed protocluster candidates that we discuss in Section~\ref{sec:halo_mass}. We used the \texttt{SubLink} merger trees \citep{rodriguez-gomez_merger_2015} to follow the descendant branches of these massive high-redshift haloes to find their eventual present-day mass. Figure~\ref{fig:Halo_mass_evolution} shows this evolution, where we interpolate the evolution of individual haloes to a finer redshift grid to fill voids in the redshift range of the TNG-Cluster and TNG300 simulations. Each hexagonal bin in Figure~\ref{fig:Halo_mass_evolution} is coloured by the median present-day mass of haloes in that bin, only when that median exceeds $M_{\rm 200c}>10^{13}\ {\rm M_{\odot}}$.

Using the uncertainty on the virial mass of \pc\ at $z=7.66$, we find the median present-day mass of haloes within that mass range to be ${\rm log_{10}}(M_{200 {\rm c},\ z=0}[{\rm M_{\odot}}]) = 14.3\pm0.6$. Assuming \pc\ follows this median evolutionary pathway, it is poised to become a low-mass cluster at $z=0$, consistent within uncertainties with a Fornax-type cluster ($M_{200 {\rm c},\ z=0} =  0.7-3 \times 10^{14}\ {\rm M_{\odot}}$). By contrast, the observed virial mass of A2744-PC-z7p9 places it on course to reach ${\rm log_{10}}(M_{200 {\rm c},\ z=0}[{\rm M_{\odot}}]) = 14.7\pm0.5$ (assuming it follows the median evolutionary pathway), consistent with Virgo-type clusters ($M_{200{\rm c},\ z=0} =  0.3 - 1 \times 10^{15}\ {\rm M_{\odot}}$). 

Notably, although high-mass haloes at $z>7$ appear to be a strong diagnostic of high virial masses at $z=0$, the reverse is not as clear. We find examples of relatively low-mass haloes at $z<7$ that grow rapidly to become clusters at the present day. As such, the average growth histories of $z=0$ clusters are steeper than the evolutionary pathways of high-mass haloes at $z>7$. 

\section{Evidence for enhanced environmental effects with increasing halo mass}
\label{sec:enh_env_eff}

There are five $z\gtrsim7$ protoclusters for which we can retrieve stellar-mass and positional information to make our virial-mass estimates. Two of these protoclusters have been claimed to host `evolved' (strong Balmer breaks and low-equivalent-width emission lines) galaxies (A2744-PC-z7p9, \citealt{witten_before_2025}; SPT0311-58, \citealt{arribas_ga-nifs_2024}) while the other three (\pc, this work; JADES-GS-OD-7.2, \citealt{Li2026}, \citealt{helton_identification_2024}; MACS0416-OD-z8p5, \citealt{fudamoto_sapphires_2025}) show evidence of highly star-forming galaxies dominated by young stellar populations. The latter are consistent with the typical properties of observed field galaxies at the same redshift \citep{roberts-borsani_between_2024}.

Interestingly, the protoclusters that show evidence of evolved stellar populations have the highest virial masses, as determined by our adapted SMHM relation: A2744-PC-z7p9 is 0.2 dex more massive than \pc, while SPT0311-58 is an order of magnitude more massive than \pc. The SFMS shown in Figure~\ref{fig:SFMS} provides early evidence for suppression of star formation in the most massive protoclusters. While the most massive galaxies ($M_{\star}\sim10^9\ {\rm M_{\odot}}$) in A2744-PC-z7p9 often fall on or below the SFMS, in the lower halo mass protoclusters (\pc\ and MACS0416-OD-z8p5) they fall above it. In this mass regime, we expect to be highly complete and able to observe galaxies even in low-SFR phases \citep{simmonds_bursting_2025}; therefore, the differing scatter from the SFMS indicates a suppression of star formation with increasing halo mass. However, we note this suppression is unlikely to lead to permanent quenching at this epoch.

\begin{figure}
\centering
\includegraphics[width=0.5\textwidth]{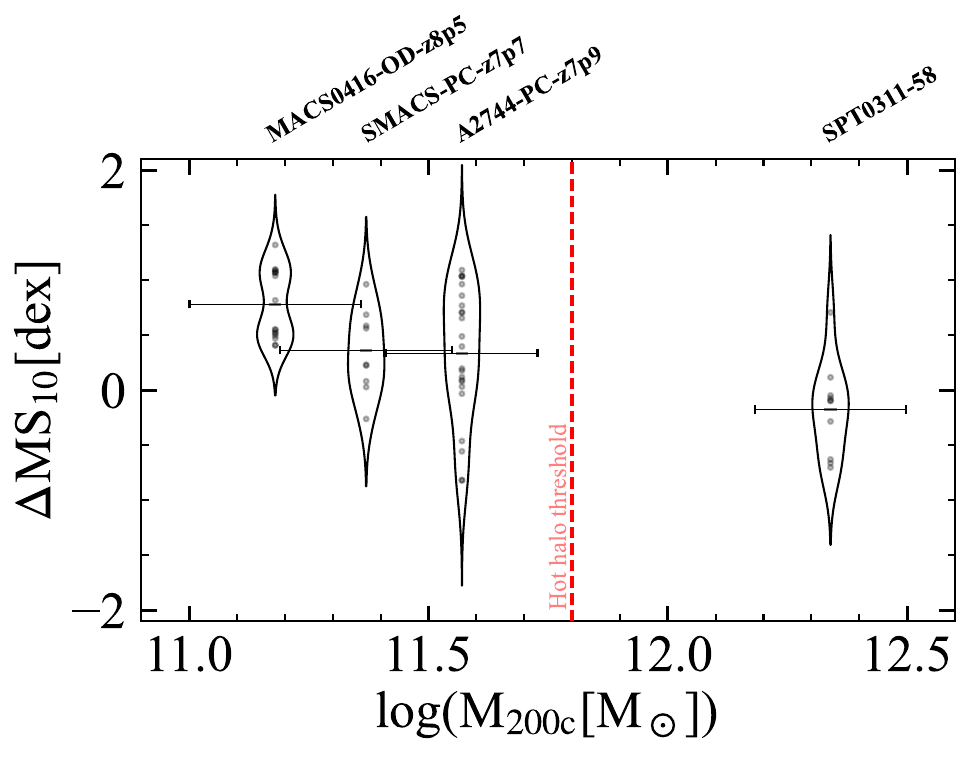}
\caption{Offset from the main sequence, $\Delta {\rm MS_{10}} = {\rm log(SFR_{10}/SFR_{MS,10})}$,  of different $z\gtrsim7$ protoclusters at their respective redshifts, as a function of virial mass. The protoclusters, ordered by virial mass are MACS0416-OD-z8p5 \citep{fudamoto_sapphires_2025}, SMACS-PC-z7p7 (this work), A2744-PC-z7p9 \citep{witten_before_2025}, and SPT0311-58 \citep{arribas_ga-nifs_2024} (with references for stellar masses and ${\rm SFR_{10}}$). The central bar indicates the average offset, and the width of the bar shows the uncertainty in the halo mass. Individual galaxies are shown as grey data points. The threshold at which a hot halo typically forms is indicated by a dashed red line \citep[e.g.][]{dekel_galaxy_2006}. A population of galaxies that fall below the SFMS emerges with increasing halo mass, resulting in a decline in the average offset. } 
\label{fig:SFMS_scatter}
\end{figure}

Figure~\ref{fig:SFMS_scatter} shows the offset from the SFMS with an averaging time-scale of 10 Myr, SFMS$_{10}$ (using the parametrisation of \citealt{mcclymont_thesan-zoom_2025}), for galaxies identified in the aforementioned protoclusters (that have SFR$_{10}$ and $M_{\star}$ measurements in the literature), as a function of protocluster halo mass. While these halo masses have large uncertainties, it is clear that with increasing halo mass there is an emergence of a population of galaxies that fall below the SFMS. We also see an evolution in the average offset of each protocluster from the SFMS, from lying above the SFMS at lower protocluster halo mass to being consistent with the SFMS at high halo masses. 

We hypothesise that the enhanced SFR in the lower virial mass protoclusters are driven by the significant gas accretion in protocluster environments \citep{fudamoto_early_2025,Heintz2026}. However, virial temperature increases with virial mass. At $M_{200 \rm c}\sim 10^{11.8}\ {\rm M_{\odot}}$, the virial temperature exceeds the temperature at which cooling is effective \citep{mo_galaxy_2010}, resulting in heating of the CGM to temperatures of $T>10^6$ K. While cold filamentary structures are dense enough at these redshifts ($z\gtrsim3$) to penetrate the hot halo and feed star formation, the net effect of the hot halo in suppressing star formation is poorly understood. Moreover, the formation of massive galaxies as these protoclusters evolve will result in enhanced stellar and AGN feedback, and prevalent galaxy-galaxy interactions \citep{marcelin_enhanced_2025} can significantly disrupt gas accretion. These feedback mechanisms together suppress star formation with increasing virial mass and appear to become especially effective at ${\rm log_{10}}(M_{200{\rm c}}[{\rm M_{\odot}}]) \gtrsim 11.5$ (see Figure~\ref{fig:SFMS_scatter}). This produces an increased fraction of galaxies below the SFMS in the most massive haloes. In contrast, galaxies farther from the central core of these protoclusters still have elevated SFR thanks to enhanced gas accretion and are less affected by these environmental effects that are more prominent in the cores of protoclusters.

\section{Conclusions}
\label{sec:conclusions}
We used recent \jwst\ NIRCam F444W grism observations of the SMACS J0723.3-7327 field to confirm the photometric candidate protocluster environment from \cite{laporte_lensed_2022}. These observations reveal six \OIII-emitting galaxies at $z=7.66$, a further three higher-confidence photometric candidates (in close proximity to these spectroscopically confirmed galaxies), and finally two $z\sim7.6$ photometric candidates. These galaxies reside within a $100 \times 200\ {\rm pkpc^2}$ area, with a redshift range of $\Delta z = 0.005$, yielding an overdensity parameter of $\delta\sim 200$. The UVLF of this small volume is several hundred times ($320\times$) the nominal UVLF and a factor of six  lower than that of the A2744-PC-z7p9 protocluster at $z=7.88$ \citep{hashimoto_reionization_2023, morishita_early_2023, witten_before_2025}. 

We assessed the properties of these galaxies using the SED-fitting code \texttt{Prospector}. \pc-resident galaxies are exclusively characterised by a recent, ongoing burst of star formation (${\rm SFR_{10}/SFR_{100}}>1$) and typically blue UV slopes for their redshift. This contrasts with the properties of galaxies in the $z=7.88$ protocluster, A2744-PC-z7p9. Galaxies in A2744-PC-z7p9 show SFHs that range from recent, ongoing bursts in the outskirts to continuous and even declining SFHs within the core of the protocluster \citep{witten_before_2025}. 

We find evidence of excessive damping around the \Lya-break, caused by extreme neutral hydrogen column densities ($N_{\rm HI} > 10^{21.5} \ {\rm cm^{-2}}$), in two \pc-resident galaxies. However, this is not widespread across our \pc-resident sample, contrasting the high column densities seen in $\sim 60 \%$ of A2744-PC-z7p9-resident galaxies \citep{witten_before_2025}. 

To estimate the halo masses of \pc\ and other high-redshift protoclusters, we derived a relation between stellar mass within the virial radius and the virial mass itself. To do so, we used the TNG-Cluster and TNG300 simulations, which extend to higher virial masses at high redshift than typical SMHM relations \citep[e.g.][]{behroozi_universemachine_2019}. With our adapted stellar within-the-virial-radius-to-halo mass relation, we estimate a virial mass for \pc\ of ${\rm log_{10}}(M_{200 \rm c} [{\rm M_{\odot}}]) = 11.4 \pm 0.2$, 0.2 dex lower than that of A2744-PC-z7p9. Assuming that \pc\ follows the average evolutionary pathway of similar mass high-redshift haloes in these simulations, it is expected to become a Fornax-type cluster at $z=0$, while A2744-PC-z7p9 appears to be a progenitor of a Virgo-type cluster. We additionally note that although high-mass (${\rm log_{10}}(M_{200 \rm c} [{\rm M_{\odot}}]) \gtrsim 11.5$), high-redshift ($z>7$) haloes are on course to become clusters at the present day, present-day clusters do not necessarily have high mass at high redshift. Instead, they can grow rapidly after beginning significant accretion at later cosmic epochs. 

The average offset from the SFMS of protocluster-resident galaxies within protoclusters appears to evolve with increasing halo mass. At low halo masses, protocluster-resident galaxies largely lie above the SFMS, presumably due to enhanced gas accretion expected in these environments. At large stellar masses ($M_{\star} > 10^9\ {\rm M_{\odot}}$), where we expect to be highly complete, massive galaxies transition from sitting above the SFMS to having suppressed SFR relative to the SFMS with increasing protocluster halo mass.

We hypothesise that this emerging star-formation suppression is driven by multiple feedback processes that are related to increasing protocluster halo mass (${\rm log_{10}}(M_{200 \rm c} [{\rm M_{\odot}}]) \gtrsim 11.5$). These high halo masses cause virial heating of the CGM gas, enhanced gas accretion, and hence stellar and AGN feedback, and high merger rates \citep[that can significantly impact the gas distribution around these core-resident galaxies, e.g.][]{witten_deciphering_2024}. The exact combination of these processes that produces more evolved galaxies is however uncertain. 

To place stronger constraints on the evolution of protocluster-resident galaxies across differing halo masses, we must increase the number of robust high-redshift protocluster candidates. With existing simulations, such as TNG-Cluster, we can envisage homogeneous selection criteria for high-redshift protoclusters. Applying these selection criteria to existing and future deep NIRCam imaging and grism campaigns will reveal a larger population of protocluster environments; targeted follow-up will reveal the properties of resident galaxies. Combining this with rapid advancements in large-scale cosmological radiative simulations, we will soon have a complementary picture of galaxy evolution in extreme environments from both observational and theoretical vantage points.

\begin{acknowledgements}
The authors would like to thank Annalisa Pillepich and William McClymont for fruitful discussion and useful comments which improved this manuscript.

The work presented in this paper is based on observations made with the NASA/ESA/CSA {\it James Webb} Space Telescope. The data were obtained from the Mikulski Archive for Space Telescopes at the Space Telescope Science Institute, which is operated by the Association of Universities for Research in Astronomy, Inc., under NASA contract NAS 5-03127 for \jwst. These observations are associated with program 2736 and 4043.

This work has received funding from the Swiss State Secretariat for Education, Research and Innovation (SERI) under contract number MB22.00072, as well as from the Swiss National Science Foundation (SNSF) through project grant 200020\_207349. 
The Cosmic Dawn Center (DAWN) is funded by the Danish National Research Foundation under grant DNRF140. 
JSB acknowledges support from the Simons Collaboration on "Learning the Universe". 
WMB gratefully acknowledges support from DARK via the DARK fellowship. This work was supported by a research grant (VIL54489) from VILLUM FONDEN. 
LRI acknowledges support by the Science and Technology Facilities Council (STFC), ERC Advanced Grant 695671 “QUENCH" and the UKRI Frontier Research grant RISEandFALL.

\end{acknowledgements}

\bibliographystyle{aa}
\bibliography{references_downloaded} 

\begin{appendix}
\section{SEDs}
\label{app:SEDs}

In this appendix, we expand on our discussion of our SED fitting of \pc\ galaxies. We first report the photometry that we measure for each galaxy in Table~\ref{tab:phot} utilising the \texttt{Photutils} package.

\begin{figure*}
    \centering
    \includegraphics[width=1\linewidth]{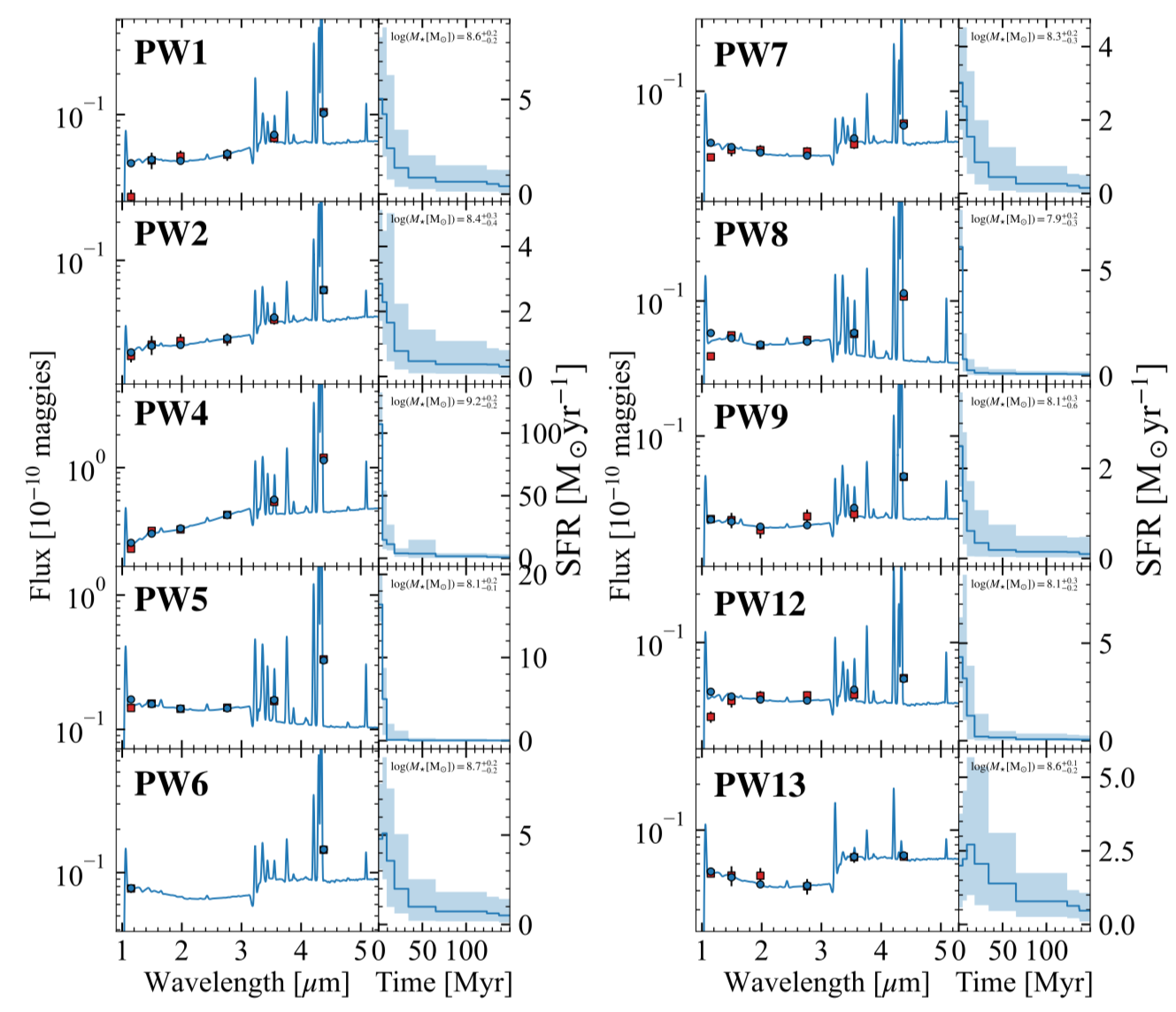}
    \caption{(Left:) The SEDs of \pc-resident galaxies (red squares) and the best fit photometry (blue circles) and spectrum (blue solid line) from our \texttt{Prospector} modelling. (Right:) The corresponding best-fit SFH (blue solid line) and associated uncertainty (blue shaded region). The best-fit stellar mass is also indicated at the top of the panel.}
    \label{fig:SEDs}
\end{figure*}

We fit this photometry, in combination with the emission line fluxes reported in Table~\ref{tab:fluxes}, with the SED-fitting code \texttt{Prospector}. We show the best-fit SEDs and SFHs from this SED fitting in Figure~\ref{fig:SEDs}. All of the galaxies in \pc\ are characterised by ongoing bursts in their SFHs. While some galaxies show evidence of mild Balmer breaks, which correspond to the galaxies with older stellar populations, the majority are characterised by having strong emission lines. The SFHs of PW4 and PW5 are consistent with those reported in \cite{tacchella_jwst_2023}.

\begin{table}
    \centering
    \caption{Emission line fluxes obtained from F444W grism observations.}
    \begin{tabular}{lccc}
        \hline
    ID & & Flux ($10^{-18}$ erg $\rm{s}^{-1}$ $\rm{cm}^{-2}$)&\\
    & $\rm{H}\beta$ & $[\ion{O}{III}] 4959$ & $[\ion{O}{III}] 5007$\\
    \hline
    PW1 &$0.6 \pm 0.2$&$0.7 \pm 0.4$&$2.0 \pm 0.4$\\
    PW3 & $<0.5$ & $<0.5$ & $1.5 \pm 0.5 $\\
    PW4 & $5.3 \pm 0.3$ & $14.6\pm 0.6$ & $43.5\pm0.6$ \\
    PW5 & $1.2 \pm 0.2$ & $3.5 \pm 0.4$ & $10.4 \pm 0.4$\\
    PW6 & $<0.6$ & $1.2 \pm 0.6$ & $3.5 \pm 0.6$\\
    PW8 & $0.6\pm 0.3$ & $2.0 \pm 0.6$ & $6.1\pm0.6$ \\
    \hline
    \end{tabular} 
    \\
    Notes: The reported fluxes are not corrected for magnification.
    \label{tab:fluxes}
\end{table}

\begin{table*}
    \centering
    \caption{Photometric measurements of protocluster-resident galaxies.}
    \begin{tabular}{lccccccc}
        \hline
    ID & F090 & F115 & F150 & F200 & F277 & F356 & F444 \\
    \hline
    PW1 & $ <2.79$
& $13.11  \pm  1.69$
& $24.43  \pm  3.40$
& $26.17  \pm  2.64$
& $26.89  \pm  2.66$
& $35.80  \pm  2.80$
& $55.69  \pm  1.38$ \\
    PW2 &   $ <1.71$
& $9.39  \pm  1.03$
& $11.55  \pm  1.98$
& $12.35  \pm  1.62$
& $12.72  \pm  1.42$
& $18.15  \pm  1.53$
& $31.14  \pm  0.91$\\
    PW3 & $ <1.14$
& $9.09  \pm  0.73$
& $10.34  \pm  1.90$
& $12.26  \pm  1.55$
& $17.27  \pm  1.67$
& $35.50  \pm  1.83$
& $100.53  \pm  0.95$\\
    PW4 &$ <3.36$
& $90.38  \pm  2.48$
& $131.94  \pm  3.80$
& $135.87  \pm  3.25$
& $185.09  \pm  2.80$
& $242.78  \pm  2.94$
& $617.51  \pm  2.05$\\
    PW5 & $ <2.48$
& $72.81  \pm  2.37$
& $78.59  \pm  2.80$
& $71.73  \pm  2.33$
& $73.21  \pm  1.98$
& $81.95  \pm  1.97$
& $167.56  \pm  1.40$\\
    PW6 & $<4.08$& $40.25  \pm  2.64$&OOF&OOF&OOF&OOF& $74.79  \pm  2.23$\\
    PW7 & $ <1.71$
& $19.66  \pm  1.04$
& $22.02  \pm  2.04$
& $21.92  \pm  1.69$
& $21.47  \pm  1.49$
& $23.93  \pm  1.60$
& $32.75  \pm  0.93$\\
    PW8 & $ <1.72$
& $20.20  \pm  1.05$
& $29.22  \pm  2.08$
& $24.47  \pm  1.71$
& $26.98  \pm  1.53$
& $29.92  \pm  1.63$
& $57.69  \pm  0.97$\\
    PW9 & $ <1.07$
& $11.96  \pm  0.73$
& $11.74  \pm  1.56$
& $9.84  \pm  1.31$
& $12.51  \pm  1.59$
& $13.00  \pm  1.64$
& $24.90  \pm  0.78$\\
    PW12 & $ <2.19$
& $20.27  \pm  1.63$
& $25.52  \pm  2.35$
& $27.37  \pm  1.97$
& $27.61  \pm  1.63$
& $27.88  \pm  1.64$
& $35.32  \pm  1.17$\\
    PW13 & $ <2.68$
& $28.03  \pm  1.52$
& $27.21  \pm  4.08$
& $27.15  \pm  3.35$
& $23.11  \pm  2.70$
& $35.88  \pm  2.87$
& $36.20  \pm  1.22$\\
    \hline
    \end{tabular}
    \\
    The fluxes reported are in nJy and calculated utilising an aperture size covering the F444W emission from each galaxy, without contamination from nearby sources. 1$\sigma$ limiting depths are indicated in the case of non-detection. OOF indicates a galaxy is outside of the field of imaging. All values included are not corrected for magnification.
    \label{tab:phot}
\end{table*}

\section{LRD diagnostics}
\label{app:LRD}

\subsection{Little Red Dot}
\label{sec:LRD}

\begin{figure}
    \centering
    \includegraphics[width=0.9\linewidth]{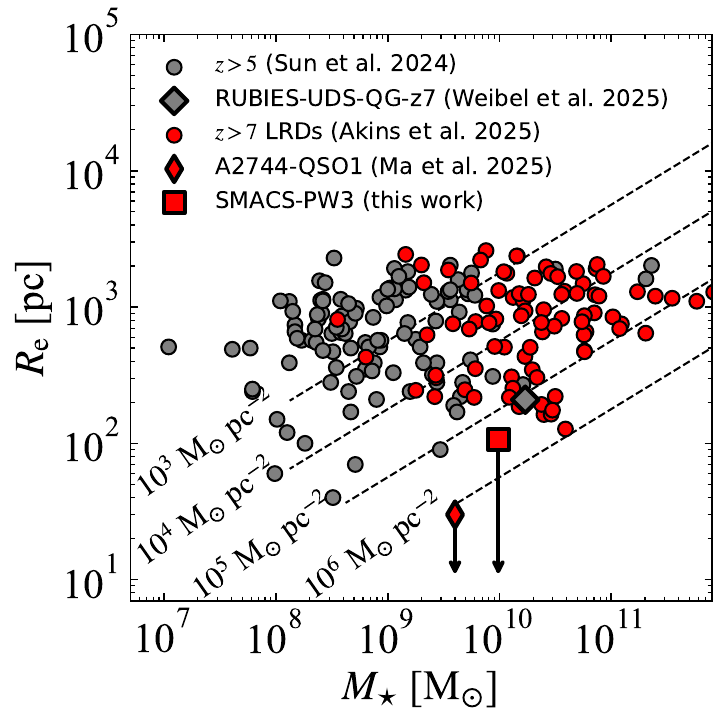}
    \caption{The stellar mass surface densities of galaxies at $z>5$ in the CEERS survey \citep[grey circle;][]{sun_structure_2024} and one of the most massive, compact, non-LRD objects (grey data points) at high-redshift \citep[grey rotated square;][]{weibel_rubies_2025}. In red we indicate a sample of photometrically identified LRDs from \citep{akins_cosmos_25} and two spectroscopically confirmed LRDs, A2744-QSO1 \citep[red diamond;][]{ma_uncover_2024} and SMACS-PW3 (red square; this work). It is clear that when modelled with purely stellar populations, the LRDs (red data points) have significantly higher stellar mass surface densities than the sample of $z>5$ galaxies. Moreover, PW3 has a higher stellar mass surface density than even one of the most extreme density objects RUBIES-UDS-QG-z7.}
    \label{fig:SMSD}
\end{figure}

\begin{figure*}
    \centering
    \includegraphics[width=0.9\linewidth]{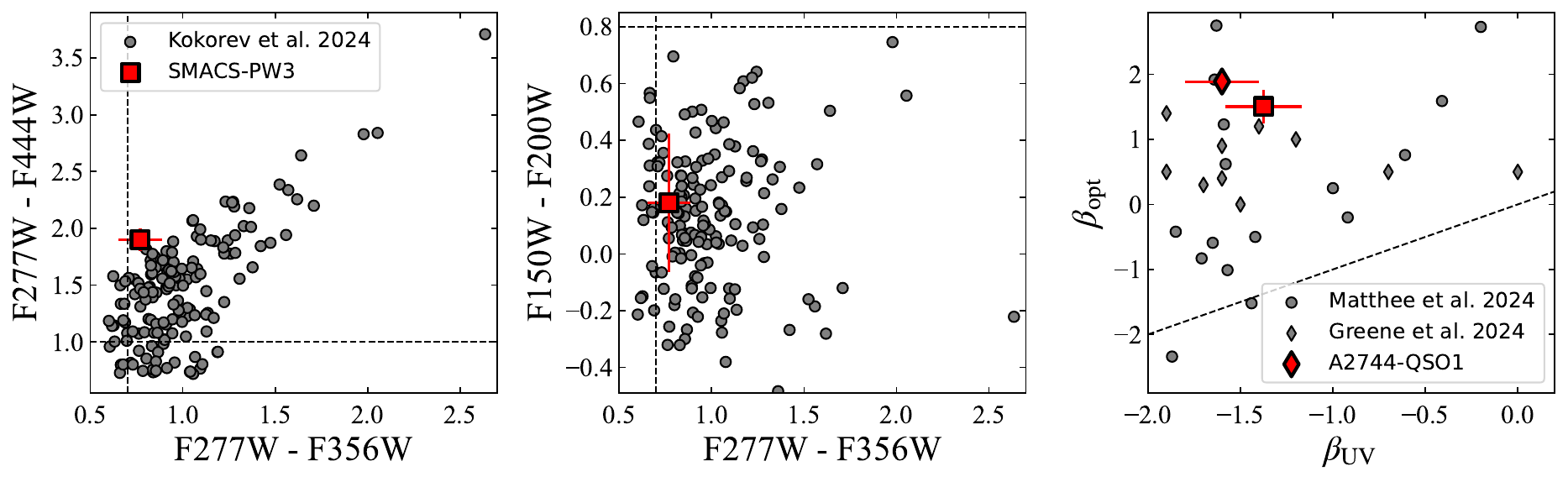}
    \caption{(Left and centre:) Colour-colour selection of LRDs utilising the selection criteria of \citet{Labbe2025} (indicated with dashed lines). The values we derive for PW3 (indicated by a square point) fall within all of the selection criteria. Examples of photometrically selected LRDs at $z>7$ from \citet{kokorev_census_2024} are indicated (circular points). (Right:) The UV and optical slopes are assumed to be two independent power laws, after correcting F444W for the emission line flux seen in the grism spectra. Two notable $z>7$ LRDs are indicated, both with very similar UV and optical slopes: PW3 from this work (red square) and QSO1 from \citet{furtak_jwst_2023} (red diamond). Spectroscopically confirmed LRDs at $z=4.2-5.5$ \citep[from][grey circles]{matthee_little_2024} and at $z>5$ \citep[from][grey diamonds]{greene_uncover_2024} are indicated. The black dashed line indicates where the UV and optical slopes are equal. We expect most LRDs to fall above this relation. It is clear that PW3 satisfies the LRD selection criteria and has typical UV and optical slopes for an LRD at $z>7$.}
    \label{fig:LRD_diagnostics}
\end{figure*}

In light of the detection of the compact, red source, PW3, we compare its photometric properties to the little red dot (LRD) selection criteria. We first fit the UV ($\sim 1-3\mu$m) and optical ($\sim 3-5\mu$m; where the emission lines fluxes have been subtracted from the F444W flux) continua with separate power laws (i.e. $f\propto \lambda^{\beta}$), revealing two distinct continuum slopes: $\beta_{\rm UV}= -1.4\pm 0.2$ and $\beta_{\rm opt}=1.5\pm 0.3$. These appear in line with the UV and optical slopes of $z\sim 5-6.5$ LRDs reported in \cite{matthee_little_2024}. Moreover, PW3 satisfies the widely utilised LRD selection criteria for galaxies at such redshifts - the ``red2'' criteria from \cite{Labbe2025}:
\begin{align*} 
&\rm F150W - F200W < 0.8 \\
&\rm F277W - F356W > 0.7 \\
&\rm F277W - F444W > 1.0\ ,
\end{align*}
which is combined with a compactness criteria: $f_{\rm F444W}(0^{\prime\prime}.4)/f_{\rm F444W}(0^{\prime\prime}.2) < 1.7$. PW3 satisfies all of the colour selection, as indicated in Figure~\ref{fig:LRD_diagnostics}. Its compactness of $f_{\rm F444W}(0^{\prime\prime}.4)/f_{\rm F444W}(0^{\prime\prime}.2) = 1.06\pm0.02$ easily satisfies the requirement.

We start with the assumption that the SED of PW3 has only stellar origins. We utilise the SED-fitting code \texttt{Prospector}, as described in Section~\ref{sec:sample}. This fitting yields a SFH that is dominated by a recent burst, over the previous $\sim 20$ Myrs, with a peak star-formation rate of $\sim 1,000 {\rm M_{\odot}/yr}$. The best-fit stellar mass is ${\rm log_{10}(}M{\rm_{\star}[M_{\odot}])}= 10 \pm 0.1$, with a V-band dust attenuation of $A_{V} = 2.0^{+0.4}_{-0.2}$.

Before discussing the failure of our \texttt{Prospector} model to reproduce the extremely red rest-optical continuum of PW3, we first consider the implied stellar mass surface density of PW3 according to our model. The radial profile of the emission from PW3 in all filters can be described by the point-source function of said filter, as such yielding an effective radius in the F115W filter of $0^{\prime\prime}.03$. At $z=7.67$ and accounting for a magnification of $\mu = 1.40$, this corresponds to $R_{\rm e}=107$ pc. This implies PW3 has a stellar mass surface density of ${\rm log_{10}(\Sigma_{*,<R_{e}}[M_{\odot}/kpc^{2}])} > 11.4\pm0.1$, as shown in Figure~\ref{fig:SMSD}. LRDs have higher stellar mass surface densities than the general galaxy population when they are SED-fit assuming only a stellar model. Relative to a sample of $z>7$ LRDs, from \cite{akins_cosmos_25}, PW3 is among the highest stellar mass surface density objects, in part due to the magnification facilitating a smaller constraint on the effective radius. Similarly, A2744-QSO1 \citep[an LRD whose rest-optical continuum has recently been shown to be driven by AGN emission;][]{furtak_high_2024, ji_blackthunder_2025, furtak_investigating_2025,deugenio_blackthunder_2025} has a significantly smaller effective radius constraint thanks to lensing magnification, and as noted by \cite{ma_uncover_2024}, this results in a stellar mass surface density in PW3 that is in excess of the maximum density observed in the local Universe and low-redshift galaxies \citep[${\rm log_{10}(\Sigma_{*,<R_{e}}[M_{\odot}/kpc^{2}])}\sim11$;][]{hopkins_maximum_2010}. Moreover, the lower bound of the stellar mass surface density of PW3 is already in excess of the highest density object observed at $z>7$ so far, RUBIES-UDS-QG-z7, a resolved, massive quiescent galaxy at $z=7.29$ \citep{weibel_rubies_2025}.

\begin{figure}
    \centering
    \includegraphics[width=0.9\linewidth]{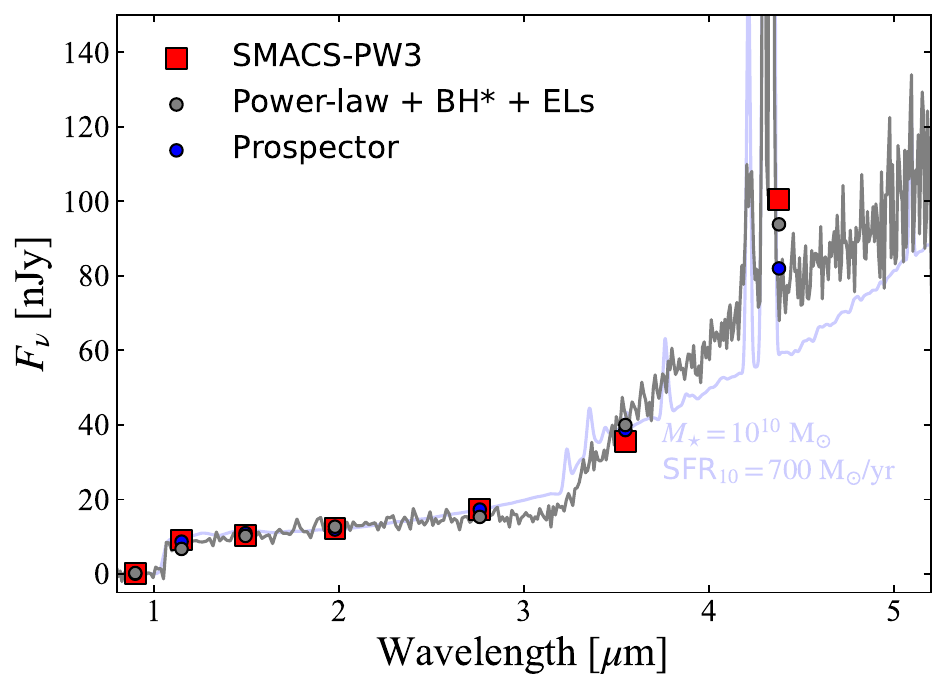}
    \caption{The photometry of the LRD candidate PW3 (red squares). We additionally show the best-fit \texttt{Prospector} model photometry (blue circles) and spectrum (blue solid line). Finally, we combine a power-law fit to the UV, with emission line constraints from the grism spectrum and the rescaled BH* spectrum from \cite{naidu_black_2025} to produce a combined spectrum (grey solid line) and photometry (grey circles). It is clear that our combined spectrum with the BH* reproduces the observed SED, while the \texttt{Prospector} model fails to match the rest-optical continuum.}
    \label{fig:LRD_models}
\end{figure}

While the \texttt{Prospector} model can fit the rest-UV-continuum well, it begins to struggle after the Balmer break (see Figure~\ref{fig:LRD_models}). As it struggles to recreate the steep slope between the F277W and F444W detections, it overpredicts the F356W emission and significantly underpredicts the F444W emission. This failure to fit the optical continuum of LRDs with only a stellar model has been noted by \cite{wang_rubies_2024}, who invoke a combination of AGN accretion disk and stellar emission to fit the rest-optical continuum. 

We note here that when we combine our best-fit UV slope model with the rescaled spectrum of the black hole star (BH*) from \cite{naidu_black_2025} and our emission line constraints (see Table~\ref{tab:fluxes}), we produce an SED that is consistent with that observed for PW3, as shown in Figure~\ref{fig:LRD_models}. Given our relative ease at producing an SED that is consistent with PW3 by invoking the BH* spectrum, and the inability of \texttt{Prospector}, even while assuming extreme stellar masses and dust attenuation, to reproduce the SED, we take this as further evidence that PW3 is an LRD.

\end{appendix}

\end{document}